\documentclass{article}
\usepackage{nips15submit_e,times}
\usepackage{hyperref}
\usepackage{url}
\usepackage{graphicx}

\usepackage{tikz}
\usepackage{ragged2e}
\usepackage{pgfplots}
\usepackage{hhline}
\usepackage{pgfmath}
\usetikzlibrary{calc}
\usetikzlibrary{shapes,arrows}
\usepackage{filecontents}
\usepackage{pgfplotstable}
\usepackage{amsmath,bm}
\usepackage{rotating}
\usepackage{multirow}
\usepackage{authblk}
\usepackage{acronym}
  \newacro{cnn}[CNN]{Convolutional Neural Network}
  \newacro{bnn}[BNN]{Binary Neural Network}
  \newacro{pe}[PE]{Processing Element}
  \newacro{simd}[SIMD]{Single Instruction on Multiple Data}
  \newacro{dsp}[DSP]{Digital Signal Processing}
  \newacro{hdl}[HDL]{Hardware Description Language}
  \newacro{le}[LE]{Logic Elements}
  \newacro{fpga}[FPGA]{Field-Programmable Gate Array}
  \newacro{fifo}[FIFO]{First-In First-Out}
  \newacro{gpu}[GPU]{Graphics Processing Unit}
  \newacro{haddoc}[HADDOC]{Hardware Automated Dataflow Description Of CNNs}
  \newacro{hls}[HLS]{High-Level Synthesis}
  \newacro{moc}[MoC]{Model of Computation}
  \newacroplural{moc}[MoC]{Models of Computation}
  \newacro{ocr}[OCR]{Optical Character Recognition}
  \newacro{qos}[QoS]{Quality of service}
  \newacroplural{qos}[QoS]{Qualities of service}
  \newacro{tpr}[TPR]{True Positive Rate}
  \newacro{mac}[MAC]{Multiply Accumulate}
  \newacro{le}[LE]{Logic Element}
  \newacro{fc}[FC]{Fully Connected}
  \newacro{simd}[SIMD]{Single Instruction on Multiple Data}
  \newacro{vhdl}[VHDL]{VHSIC Hardware Description Language}
  \newacro{lut}[LUT]{Look-Up Table}
  \newacro{nef}[NEF]{Neighborhood Extraction Factorization}  
  \newacro{ne}[NE]{Neighborhood Extraction} 
  \newacro{hdl}[HDL]{Hardware Description Language}
  \newacro{rtl}[RTL]{Register Transfer Level}
  \newacro{ip}[IP]{Intellectual Property}
  \newacro{dhm}[DHM]{Direct Hardware Mapping}
  \newacro{dag}[DAG]{Direct Acyclic Graph}
  \newacro{sdfg}[SDFG]{Synchronous DataFlow Graph}
  \newacroplural{ip}[IP]{Intellectual Properties}

\definecolor{ol_green}{rgb}{0.0, 0.5, 0.0}
\newcommand{\added}[1]{\textcolor{black}{#1}}

\pgfplotsset{compat=1.14}

\begin{document}
\iclrfinalcopy
\title{Tactics to Directly Map CNN Graphs on\\ Embedded FPGAs}

\author[1]{Kamel Abdelouahab}
\author[1,2]{Maxime Pelcat}
\author[1]{Jocelyn S{\'e}rot}
\author[3]{C{\'e}dric Bourrasset}
\author[1]{Fran\c{c}ois Berry}

\affil[1]{Institut Pascal, Univ{\'e}rsite Clermont Auvergne, France }
\affil[2]{IETR, INSA Rennes, France}
\affil[2]{Atos/Bull CEPP CINES, France}
\date{}                     
\setcounter{Maxaffil}{0}
\renewcommand\Affilfont{\itshape\small}

\maketitle

\begin{abstract}
Deep \acp{cnn} are the state-of-the-art in image classification. Since \ac{cnn} feed forward propagation involves highly regular parallel computation, it benefits from a significant speed-up when running on fine grain parallel programmable logic devices. As a consequence, several studies have proposed FPGA-based accelerators for \acp{cnn}. However, because of the large computational power required by \acp{cnn}, none of the previous studies has proposed a \emph{direct} mapping of the \ac{cnn} onto the physical resources of an FPGA, allocating each processing actor to its own hardware instance. 

In this paper, we demonstrate the feasibility of the so called \emph{direct hardware mapping (DHM)} and discuss several tactics we explore to make DHM usable in practice. As a proof of concept, we introduce the \textsc{haddoc2} open source tool, that automatically transforms a \ac{cnn} description into a synthesizable hardware description with platform-independent direct hardware mapping\footnote{This is a pre-print version. Please refer to the original paper in~\cite{Abdelouahab2017}}.
\end{abstract}


\section {Introduction}
Convolutional Neural Networks (CNNs)~\cite{Lecun1998} have become a {\em{de-facto}} standard for increasing the robustness and accuracy of machine vision systems. However, this accuracy comes at the price of a high computational cost. 
As a result, implementing \acp{cnn} on embedded devices with real-time constraints is a challenge. A solution to this challenge is to take advantage of the massive fine grain parallelism offered by embedded \acp{fpga} and benefit from the extensive concurrency exhibited by \ac{cnn}-based algorithms. By embedded \acp{fpga}, we refer to devices with limited power consumption and cost, typically under 20W and 300\$. When porting a \ac{cnn} to an embedded \ac{fpga}, the problem boils down to finding an efficient mapping between the computational model of the \ac{cnn} and the execution model supported by the \ac{fpga}. Based on works related to the implementation of real-time vision applications on FPGA-powered embedded platforms~\cite{Serot2016}, we advocate the use of a \emph{dataflow} model to solve this mapping problem. In this approach, a \ac{cnn} algorithm is described as a graph of dataflow actors exchanging data through unidirectional channels and this dataflow graph is statically and physically mapped onto the target FPGA using a library of pre-defined computing elements implementing actors. 

In the sequel,  we demonstrate the feasibility of the \emph{\ac{dhm}} approach for implementing CNN-based applications onto embedded \acp{fpga}. \ac{dhm} associates each \ac{cnn} processing entity to private resources, maximizing parallelism. To support this demonstration, we introduce \textsc{haddoc2}, a framework that provides a fully automated hardware generation for \acp{cnn} using \ac{dhm}. The \textsc{haddoc2} tool is compatible with the Caffe deep learning framework~\cite{Jia2014} and generates platform-independent VHDL synthesizable code.

The paper is organized as follows. Section~\ref{sec:related-work} reviews state-of-the-art implementations of \acp{cnn} on \acp{fpga}. Section~\ref{sec:overview} recalls the main features of \acp{cnn} from a computational point of view, focusing on parallelism issues. Section~\ref{sec:dhm} describes the \ac{dhm} approach and how it is supported by the \textsc{Haddoc2} framework. Section~\ref{sec:res} presents an assessment of the efficiency of the approach, reporting  performance and resource utilization of \acp{dhm}-based implementations for three \acp{cnn}, and Section~\ref{sec:concl} concludes the paper.

\section{Related Work} \label{sec:related-work}
Several studies leverage on FPGA computational power to implement the feed-forward propagation of \acp{cnn}. A complete review of these studies can be found in~\cite{Lacey2016}. In most approaches, CNN-based applications are implemented by mapping a limited subset of processing elements onto the target device, multiplexing in time the processing elements {and processing data in an SIMD fashion. This is the case for instance in~\cite{Qiu2016} where authors describe a \ac{cnn} accelerator implemented on a Zynq XC706 board}. 

The dataflow-based implementation of \acp{cnn} is investigated in~\cite{Farabet2012} where authors describe Neuflow, an acceleration engine for \acp{cnn} relying on a dataflow execution model.  The \ac{cnn} graph is transformed into a set of dataflow instructions, where each instruction is described as a hardware configuration of 2D-processing elements called \emph{Processing tiles (PTs)}. The execution of the graph is carried out by sequencing the instructions on an FPGA. 

The previously evoked approaches require an external memory to store intermediate results, which in turn, even with the help of a DMA, limits the final speedup. The study in~\cite{Venieris2016} features a partitioning of the \ac{cnn} graph with one bit-stream per subgraph in a way that only on-chip memory is needed to store intermediate results. This however requires the reconfiguration of the FPGA whenever data has to enter a different subgraph, which adds a substantial reconfiguration time overhead. By contrast, the \ac{dhm} approach introduced in the present paper performs all processing \emph{on the fly} and does not require any external memory to store intermediate results. Throughput is therefore not influenced by off-chip memory bandwidth.

\section{\ac{cnn} Computation} \label{sec:overview}
A typical \ac{cnn} structure 
performs a succession of convolutions interspersed with sub-sampling layers. The last layers of a \ac{cnn} are fully connected layers performing classification. Convolutional layers are the most computationally intensive layers and are commonly responsible for more than 90\% of the \ac{cnn} execution time~\cite{Cong2014}. As a consequence, we focus in this paper on the implementation of convolutional layers.

A convolutional layer $(l)$ extracts $N$ feature maps from $C$ input channels by performing $N$ convolutions of size $K \times K$ on each input. This filtering is followed by the application of a non-linear activation function $act$ and a bias term $b_n$ to each set of features. As shown in equation~\ref{convLayerProc}, $N \times C$ convolutions (resulting in $N \times C \times K \times K $ multiplications) 
are required to process a given layer. 
\begin{align}
    \label{convLayerProc}
    \forall l = &1:L   \nonumber \text{\textit{ (Number of layers)}}\\
    \forall n &=1:N^{(l)} \nonumber \text{\textit{  (Number of output feature maps)}}\\
    & \bm{f_n}^{(l)} = \mbox{act} \left[ \bm{b_n}^{(l)} + \sum_{c=1}^{C^{(l)}} conv(\bm{\phi_c^{(l)}}, \bm{w_{nc}^{(l)}}) \right]
\end{align}
where
    $\bm{f_n}^{(l)}$ is the $n^{th}$ output feature map of layer  $(l)$,
    $\bm{\phi_c^{(l)}}$ is the $c^{th}$ input feature map and
    $\bm{w_{nc}^{(l)}}$ is a pre-learned filter.

The computation described in Equation~\ref{convLayerProc} exhibits four sources of concurrency. First, \acp{cnn} have a feed-forward hierarchical structure consisting of a succession of data-dependent layers. Layers can therefore be executed in a \emph{pipelined} fashion by launching layer $(l)$ before ending the execution of layer $(l-1)$. Second, each neuron of a layer can be executed independently from the others, meaning that each of the $N^{(l)}$ element of equation~\ref{convLayerProc} can be computed in parallel. Third, all of the convolutions performed by a single neuron can also be evaluated simultaneously by computing concurrently the $C^{(l)}$ elements of equation~\ref{convLayerProc}. Finally, each 2D image convolution can be implemented in a pipelined fashion~\cite{Shoup1994} computing the $K \times K$ multiplications concurrently.

\section{Direct Hardware Mapping of CNNs} \label{sec:dhm}

A \ac{cnn} can be modeled by a dataflow process network (DPN) where nodes correspond to processing actors and edges correspond to communication channels. Each actor follows a purely data-driven execution model where execution (firing) is triggered by the availability of input operands~\cite{Lee1995}. The \ac{dhm} approach consists of \emph{physically} mapping the whole graph of actors onto the target device. Each actor then becomes a computing unit with its specific instance on the \ac{fpga} and each edge becomes a signal.  

\begin{figure}[!ht]
\centering
\includegraphics[width=0.55\textwidth]{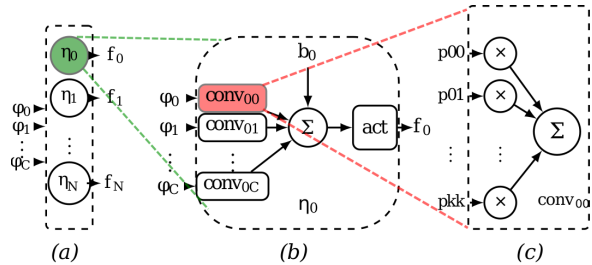}
\caption{The 3 levels of \ac{dhm} use on \ac{cnn} entities: (a) in the convolution layers, (b) in the neurons, (c) in the convolution engines.}
\label{dhm_entities}
\end{figure}

\begin{figure}[!ht]
    \centering
    \begin{tikzpicture}[scale=0.75, every node/.style={scale=0.75}]

    \tikzset{neuron/.style={draw,circle,minimum size=1.2cm}};
    \tikzset{conv/.style={draw,rectangle,rounded corners=3pt,minimum size=0.4cm}};
    \tikzset{sum/.style={draw,circle,minimum size=1cm}};
    \tikzset{act/.style={draw,rectangle,rounded corners=3pt,minimum size=1cm}};
    \tikzset{neuronBox/.style={draw,rectangle,dashed,rounded corners=10pt,minimum width=5.5cm,minimum height=1.8cm}};
    \tikzset{void/.style={}}

    \node[void]     (i0)      at(-3,2)   {$\phi^{{(C1)}}_2$};
    \node[void]     (i1)      at(-3,4)   {$\phi^{{(C1)}}_1$};
    \node[void]     (i2)      at(-3,6)   {$\phi^{{(C1)}}_0$};

    \node [conv] (ce0)  at (0,-0.5)   {{$\mbox{conv}_{42}$}};
    \node [conv] (ce1)  at (0,0)      {{$\mbox{conv}_{41}$}};
    \node [conv] (ce2)  at (0,0.5)    {{$\mbox{conv}_{40}$}};
    \node [conv] (ce3)  at (0,1.5)    {{$\mbox{conv}_{32}$}};
    \node [conv] (ce4)  at (0,2)      {{$\mbox{conv}_{31}$}};
    \node [conv] (ce5)  at (0,2.5)    {{$\mbox{conv}_{30}$}};
    \node [conv] (ce6)  at (0,3.5)    {{$\mbox{conv}_{22}$}};
    \node [conv] (ce7)  at (0,4)      {{$\mbox{conv}_{21}$}};
    \node [conv] (ce8)  at (0,4.5)    {{$\mbox{conv}_{20}$}};
    \node [conv] (ce9)  at (0,5.5)    {{$\mbox{conv}_{12}$}};
    \node [conv] (ce10) at (0,6)      {{$\mbox{conv}_{11}$}};
    \node [conv] (ce11) at (0,6.5)    {{$\mbox{conv}_{10}$}};
	\node [conv] (ce12) at (0,7.5)    {{$\mbox{conv}_{02}$}};
    \node [conv] (ce13) at (0,8)      {{$\mbox{conv}_{01}$}};
    \node [conv] (ce14) at (0,8.5)    {{$\mbox{conv}_{00}$}};
    
    \node[sum]   (sum0)      at(2,0)    {$\Sigma_{4}$};
    \node[sum]   (sum1)      at(2,2)    {$\Sigma_{3}$};
    \node[sum]   (sum2)      at(2,4)    {$\Sigma_{2}$};
    \node[sum]   (sum3)      at(2,6)    {$\Sigma_{1}$};
    \node[sum]   (sum4)      at(2,8)    {$\Sigma_{0}$};
    
    \node[act]   (act0)      at(3.5,0)    {$\mbox{act}_{4}$};
    \node[act]   (act1)      at(3.5,2)    {$\mbox{act}_{3}$};
    \node[act]   (act2)      at(3.5,4)    {$\mbox{act}_{2}$};
    \node[act]   (act3)      at(3.5,6)    {$\mbox{act}_{1}$};
    \node[act]   (act4)      at(3.5,8)    {$\mbox{act}_{0}$};

    \node[void]     (o0)      at(5.5,0)  {$f^{{(C1)}}_4$};
    \node[void]     (o1)      at(5.5,2)  {$f^{{(C1)}}_3$};
    \node[void]     (o2)      at(5.5,4)  {$f^{{(C1)}}_2$};
    \node[void]     (o3)      at(5.5,6)  {$f^{{(C1)}}_1$};
    \node[void]     (o4)      at(5.5,8)  {$f^{{(C1)}}_0$};
    
    \node[neuronBox]   (n0)      at(1.8,0)  {};
    \node[neuronBox]   (n1)      at(1.8,2)  {};
    \node[neuronBox]   (n2)      at(1.8,4)  {};
    \node[neuronBox]   (n3)      at(1.8,6)  {};
    \node[neuronBox]   (n4)      at(1.8,8)  {};
    
    \draw[->,>=latex] (i0)--(ce0.west);
    \draw[->,>=latex] (i0)--(ce3.west);
    \draw[->,>=latex] (i0)--(ce6.west);
    \draw[->,>=latex] (i0)--(ce9.west);
    \draw[->,>=latex] (i0)--(ce12.west);

    \draw[->,>=latex] (i1)--(ce1.west);
    \draw[->,>=latex] (i1)--(ce4.west);
    \draw[->,>=latex] (i1)--(ce7.west);
    \draw[->,>=latex] (i1)--(ce10.west);
    \draw[->,>=latex] (i1)--(ce13.west);

    \draw[->,>=latex] (i2)--(ce2.west);
    \draw[->,>=latex] (i2)--(ce5.west);
    \draw[->,>=latex] (i2)--(ce8.west);
    \draw[->,>=latex] (i2)--(ce11.west);
    \draw[->,>=latex] (i2)--(ce14.west);

    \draw[->,>=latex] (ce0) --(sum0)   ;
    \draw[->,>=latex] (ce3) --(sum1)   ;
    \draw[->,>=latex] (ce6) --(sum2)   ;
    \draw[->,>=latex] (ce9) --(sum3)   ;
    \draw[->,>=latex] (ce12)--(sum4)   ;
    \draw[->,>=latex] (ce1) --(sum0)   ;
    \draw[->,>=latex] (ce4) --(sum1)   ;
    \draw[->,>=latex] (ce7) --(sum2)   ;
    \draw[->,>=latex] (ce10)--(sum3)   ;
    \draw[->,>=latex] (ce13)--(sum4)   ;
    \draw[->,>=latex] (ce2) --(sum0)   ;
    \draw[->,>=latex] (ce5) --(sum1)   ;
    \draw[->,>=latex] (ce8) --(sum2)   ;
    \draw[->,>=latex] (ce11)--(sum3)   ;
    \draw[->,>=latex] (ce14)--(sum4)   ;

    \draw[->,>=latex] (sum0)--(act0);
    \draw[->,>=latex] (sum1)--(act1);
    \draw[->,>=latex] (sum2)--(act2);
    \draw[->,>=latex] (sum3)--(act3);
    \draw[->,>=latex] (sum4)--(act4);

    \draw[->,>=latex] (act0)--(o0);
    \draw[->,>=latex] (act1)--(o1);
    \draw[->,>=latex] (act2)--(o2);
    \draw[->,>=latex] (act3)--(o3);
    \draw[->,>=latex] (act4)--(o4);
    
\end{tikzpicture} 
    \caption{Applying the 3 levels of DHM (Fig.~\ref{dhm_entities}) to a convolutional layer C1 (N=5, C=3, K=3): 15 separate convolution engines (135 Multipliers and 15 adders) plus 5 adders and 5 activation blocks are required to process the fully parallel layer (bias omitted).}
    \label{dhm_layer}
\end{figure}
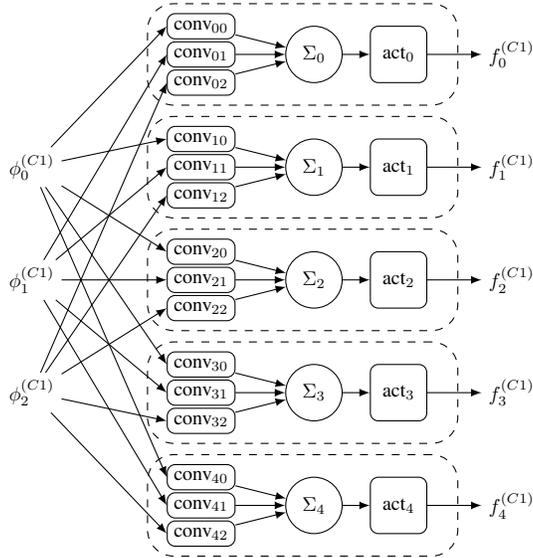

This approach fully exploits \ac{cnn} concurrency. All neurons in a layer are mapped on the device to take advantage of inter-neuron parallelism (Fig.~\ref{dhm_entities}-a). In neurons, each convolution is mapped separately (Fig.~\ref{dhm_entities}-b) and finally, within a convolution engine, each multiplier is instantiated separately (Fig~\ref{dhm_entities}-c). As an example, Fig.~\ref{dhm_layer} illustrates how a convolution layer C1 ($C=3, N=5, K=3$) extracts 5 features from a 3-feature input pixel flow. In this example, 15 convolutions and 5 activation blocks are mapped onto the \ac{fpga} as a result of the layer graph transformation, which corresponds to 135 multiplications, 20 sums and 5 activations. \added{DHM of pooling layers is also performed but lowest-level implementation elements are kept out of the scope of this paper.}

The \emph{direct hardware mapping} approach exemplified above makes external memory accesses unnecessary, while classical FPGA implementations store intermediate results or parameters on external memory. The processing is then performed \emph{on-the-fly} on \emph{streams} of feature maps. Moreover, due to the fully pipelined execution model, the global throughput is only limited by the maximum clock frequency. 

These advantages come at the cost of a high resource consumption since the whole graph has to be mapped onto the physical resources of the \ac{fpga}. This resource consumption could make \ac{dhm} impractical. It is therefore crucial for \ac{dhm} to explore tactics that efficiently translate CNN actors into hardware. The most important issues to solve are those related to the representation of numbers and the implementation of multiplications.


\subsection{Approximate fixed-point data representations}
Several studies have demonstrated that \acp{cnn}, and more generally deep learning applications, usually tolerate approximate computing with short fixed-point arithmetic. Frameworks such as Ristretto~\cite{Gysel2016} fine-tune a \ac{cnn} data representation to support fixed-point numerical representations with variable data lengths. The \ac{dhm} approach advocated in this paper takes advantage of data and parameter quantization to reduce the amount of hardware resources by first inferring the minimal required precision and then \emph{deriving} the hardware resources that exactly match this precision.

\subsection{Implementing Multiplications with Logic Elements}
\label{subsec:le_arith}

Convolutions require many multiplications. If these multiplications are implemented using hardwired \ac{dsp} blocks within the target \ac{fpga}, they become the bottleneck limiting the size of the implemented CNN. For instance, the second layer of the LeNet5 network~\cite{Lecun1998} ($C=6, N=16, K=5$) requires $2400$ multipliers, exceeding the number of multipliers provided by the DSP blocks of most FPGAs, and especially of embedded FPGAs. We overcome this problem by systematically forcing the synthesis tool to implement multiplications with logical elements instead of DSP blocks, leading the resulting implementations to rely on AND gates and trees of half-adders~\cite{Altera04}.

In addition, we take advantage of the fact that the convolution kernels -- and hence one operand of each multiplication -- are constants derived from an offline training stage. Multipliers can thus be specialized to their constants. While this approach limits the flexibility of the system because it requires to re-synthesize the VHDL design whenever parameters values are changed, it delegates to the synthesis tool the task to perform low-level area and performance optimization. More particularly, multiplications by 0 (\textit{resp} 1) are removed (\textit{resp.} replaced by a simple signal connection) and multiplications by a power of 2 are transformed into shift registers.

\subsection{Automated Hardware Generation with Haddoc2}

The \textsc{Haddoc2} framework is a set of tools built upon the \ac{dhm} principle and upon the optimization tactics described in previous section. It generates a platform-independent hardware description of a \ac{cnn} from a Caffe model~\cite{Jia2014}.
\ac{cnn} layers in \textsc{Haddoc}2 are described using a small number of basic predefined actors written in structural VHDL. These actors follow a dataflow execution semantics. The output 
can be synthesized for any FPGA device with tools supporting VHDL~93. The \textsc{Haddoc2} framework and the library of \ac{cnn} IP-cores supporting the \ac{dhm} approach are open-source and available\footnote{https://github.com/KamelAbdelouahab/haddoc2.}.

\section{Experimental Results with Haddoc2} \label{sec:res}

\begin{figure}[!ht]
    \centering
    \pgfplotstableread[row sep=\\,col sep=&]{
nbits   & lenet5    & cifar10   & svhn  \\
3       & 98.06     & 09.99     & 19.58 \\
4       & 98.82     & 24.76     & 25.55 \\
5       & 98.90     & 59.79     & 79.68 \\
6       & 98.91     & 72.48     & 85.60 \\
7       & 98.92     & 74.57     & 87.15 \\
8       & 98.95     & 75.88     & 87.35 \\
}\tprdata

\begin{tikzpicture} [scale=0.8]
\begin{axis}[
    /pgf/number format/.cd,
        1000 sep={},
	xmajorgrids,
	xminorgrids,
 	ymajorgrids,
 	yminorgrids,
    ylabel={\emph{acc} (\%)},
    xlabel={Bit-width (bits)},
    legend style={at={(1,0)},anchor=south east}
]
\addplot +[thick, const plot mark left] table[x=nbits,y=lenet5]  {\tprdata};
\addplot +[thick, const plot mark left] table[x=nbits,y=cifar10] {\tprdata};
\addplot +[thick, const plot mark left] table[x=nbits,y=svhn]    {\tprdata};
\addlegendentry{LeNet5}
\addlegendentry{CIFAR10}
\addlegendentry{SVHN}

\addplot[dashed,color=blue] coordinates {(3,98.96) (8,98.96)};
\addplot[dashed,color=red]  coordinates {(3,76.63) (8,76.63)};
\addplot[dashed,color=green!20!black]  coordinates {(3,87.54) (8,87.54)};

\end{axis}

\end{tikzpicture} 
    \caption{Evolution of classification accuracy vs bit-width for the studied CNNs. The dashed lines refers to accuracy of the baseline 32-bits floating point model.}
    \label{acc_vs_nbits}
\end{figure}
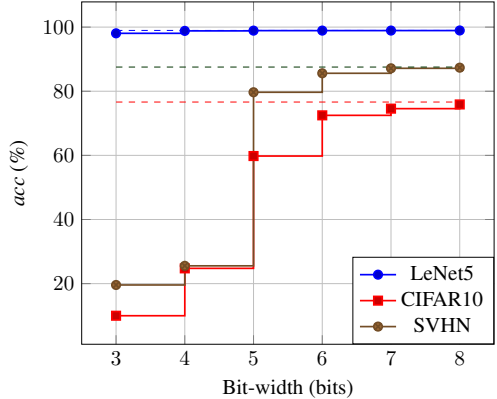

As proofs of concept, FPGA-based accelerators for three benchmark CNNs are implemented with \textsc{Haddoc2}: LeNet5~\cite{Lecun1998}, SVHN~\cite{Netzer2011} and CIFAR10~\cite{Krizhevsky2009}. Table~\ref{res:setup} details the topology of these CNNs where \emph{mpool} refers to the pooling layer that reduces the dimensionality of each feature map and \emph{tanh} is the hyperbolic tangent activation function. The Cifar10 and SVHN \acp{cnn} share the same topology with different kernel values, which is useful to study the impact of kernel proportions on a \ac{dhm}-based implementation. For each network, the fixed-point representation is chosen to respect the classification accuracy, as a result of an exploration shown in Fig.~\ref{acc_vs_nbits}. The study of quantization effects on \acp{cnn} is beyond the scope of this paper and can be found, for instance, in~\cite{Suyog2015,Gysel2016}. In our case, a 3-bit representation is chosen for the LeNet5 network and a 6-bit representation for SVHN and CIFAR10\footnotemark[2]. The shares of its zero-valued parameters, one-valued parameters and power-of-two-valued parameters are evaluated and reported in table~\ref{res:setup}. They represent, by far, more than 90\% of the parameters in all cases.
\footnotetext[2]{Similarly to~\cite{Gysel2016}, a fine tuning of the CNN parameters has been performed after selecting the bit-width, which increases the classification accuracy of the quantized CNN.}

\small
\begin{table}[!ht]
\centering
\caption{Topology of the convolutional layers of the studied \acp{cnn}.}
\begin{tabular}{l|ccc|ccc|ccc|}
\cline{2-10}
                                        & \multicolumn{3}{|c|}{LeNet5~\cite{Lecun1998}}  & \multicolumn{3}{|c|}{Cifar10~\cite{Krizhevsky2009} } & \multicolumn{3}{|c|}{SVHN~\cite{Netzer2011}}    \\ \hline
\multicolumn{1}{|l|}{Input Patches}     & \multicolumn{3}{c|}{28 x 28}       & \multicolumn{3}{c|}{32 x 32 x 3}      & \multicolumn{3}{c|}{32 x 32 x3}      \\ \hline \hline
\multicolumn{1}{|l|}{Layer parameters}  & $N$         & $C$       & $K$      & $N$         & $C$         & $K$       & $N$         & $C$         & $K$      \\ \hline 
\multicolumn{1}{|l|}{conv1+mpool+tanh}  & $20$        & $1$       & $5$      & $32$        & $3$         & $5$       & $32$        & $3$         & $5$      \\ \hline
\multicolumn{1}{|l|}{conv2+mpool+tanh}  & $50$        & $20$      & $5$      & $32$        & $32$        & $5$       & $32$        & $32$        & $5$      \\ \hline
\multicolumn{1}{|l|}{conv3+mpool+tanh}  & $-$         & $-$       & $-$      & $64$        & $32$        & $5$       & $64$        & $32$        & $5$      \\ \hline \hline
\multicolumn{1}{|l|}{accuracy float (\%)}& \multicolumn{3}{c|}{$98.96$}        & \multicolumn{3}{c|}{$76.63$}          & \multicolumn{3}{c|}{$87.54$}  \\ \hline
\multicolumn{1}{|l|}{selected bit-width} & \multicolumn{3}{c|}{$3$}            & \multicolumn{3}{c|}{$6$}              & \multicolumn{3}{c|}{$6$}      \\ \hline
\multicolumn{1}{|l|}{acc. bit-width (\%)}& \multicolumn{3}{c|}{$98.32$}        & \multicolumn{3}{c|}{$73.05$}          & \multicolumn{3}{c|}{$86.03$}  \\ \hline \hline
\multicolumn{1}{|l|}{zero parameters(\%)}& \multicolumn{3}{c|}{$88.59$}        & \multicolumn{3}{c|}{$33.78$}          & \multicolumn{3}{c|}{$37.14$}  \\ \hline 
\multicolumn{1}{|l|}{one parameters(\%)} & \multicolumn{3}{c|}{$6.31$}         & \multicolumn{3}{c|}{$45.32$}          & \multicolumn{3}{c|}{$46.50$}  \\ \hline 
\multicolumn{1}{|l|}{pow2 parameters(\%)}& \multicolumn{3}{c|}{$0.05$}         & \multicolumn{3}{c|}{$16.40$}          & \multicolumn{3}{c|}{$13.62$}  \\ \hline 
\multicolumn{1}{|l|}{other (\%)}         & \multicolumn{3}{c|}{$5.05$}         & \multicolumn{3}{c|}{$4.50$}           & \multicolumn{3}{c|}{$2.74$}  \\ \hline 
\end{tabular}
\label{res:setup}
\end{table}
In order to illustrate the impact of the developed tactics, Table~\ref{res:fit1} reports post-fitting results of a LeNet5 accelerator with a 5-bit precision on an embedded Intel Cyclone V 5CGXFC9E7 device, using 3 implementation strategies. In the first result, only DSP blocks are used to map all CNN multiplications. The resulting hardware requires $72\times$ the available resources of the device. The second case features an implementation of multiplication based on logic elements and requires $3.8\times$ the available logic. Using tailored multipliers reduces resources by a factor of $8.6\times$, fitting the CNN accelerator onto an Intel Cyclone V embedded FPGA. 

Tables~\ref{res:fit2}-a and ~\ref{res:fit2}-b respectively detail post-fitting results on two embedded \ac{fpga} platforms: the Intel Cyclone V 5CGXFC9E7 and the Xilinx Kintex7 XC7Z045FBG (using respectively Intel Quartus 16.1 and Xilinx Vivaldo 2016.4 synthetizers). To the best of our knowledge, these numbers are the first to demonstrate the applicability of a DHM-based approach for the implementation of \acp{cnn} on embedded FPGAs. The three hardware accelerators fit onto the embedded devices with no off-chip memory requirement. The reported memory footprint corresponds to line buffers used by dataflow-based convolution engines~\cite{Shoup1994} and both synthesis tools instantiate LUT-based memory blocks to implement these buffers. As expected when using \ac{dhm}, the logic utilization in the \ac{fpga} grows with the size of the \ac{cnn}. In addition, the proportion of null kernels affects the amount of logic needed to map a CNN graph. 

Finally, table~\ref{res:compare} compares Haddoc2 performance to implementations on FPGA, GPU and ASIC. For the Cifar10 CNN, we find that a direct hardware mapping approach grants  $\times2.63$ higher throughput on the same device when compared to fpgaConvNet, the state-of-the-art framework for mapping CNNs on FPGAs. For LeNet5, a $\times1.28$ acceleration is reported which corresponds to a classification rate of 64.42 HD images/sec with a 3-scale pyramid. The GPU platform delivers the best performance in terms of computational throughput but the price is a high power consumption while ASIC technology gives the best throughput per Watt trade-off at the price of lower reconfigurability and higher production costs. \added{For deeper \ac{cnn} implementations, such as in~\cite{Qiu2016}, \ac{dhm} is infeasible on current embedded FPGAs because the Logic Elements required to derive the accelerators exceed the available hardware resources.}

\added{However, and given the recent improvements of \acp{bnn} --reported for instance in FINN~\cite{Umuroglu2017}--, the implementation of deeper CNNs can be addressed by leveraging on \acp{bnn}. \acp{bnn} involve a rescheduling of the \ac{cnn} graph as well as a retraining the network to perform operations using a single bit.}
\footnotetext[1]{Performance of the feature extractor}





\begin{table}[h]
\centering
\caption{Resource utilization by a \ac{dhm} LeNet5 CNN \newline with different implementations strategies for multipliers.}
\begin{tabular}{c|c|c|c|}
\cline{2-4}
\cline{2-4}
                                        & DSP-based       & LE-based       & LE-based + const. \\ \hline
\multicolumn{1}{|l|}{Logic Usage (ALM)} & NA              & 433500 (381\%) & 50452 (44\%)     \\ \hline
\multicolumn{1}{|l|}{DSP Block usage}   & 24480 (7159 \%) & 0 (0\%)        & 0 (0\%)          \\ \hline                    
\end{tabular}
\label{res:fit1}
\end{table}

\begin{table}[h]
\centering
\caption{Resource Utilization of the three hardware accelerators:  a- an Intel Cyclone V FPGA,  b- a Xilinx Kintex 7 FPGA.}
\begin{tabular}{ll|c|c|c|}
\cline{3-5}
                                        &                           & LeNet5~\cite{Lecun1998}& Cifar10~\cite{Krizhevsky2009}& SVHN~\cite{Netzer2011}\\ \hline
\multicolumn{1}{|l|}{\multirow{5}{*}{a}}& Logic Elements (ALMs)     & 8067  (7\%)          & 51276 (45\%)         &  39513 (35\%) \\ \cline{2-5} 
\multicolumn{1}{|l|}{}                  & DSP Blocks                & 0     (0 \%)         & 0     (0\%)          &  0     (0\%)  \\ \cline{2-5} 
\multicolumn{1}{|l|}{}                  & Block Memory Bits         & 176   (1\%)          & 15808 (1\%)          &  10878 (1\%)  \\ \cline{2-5}
\multicolumn{1}{|l|}{}                  & Frequency                 & 65.71  MHz           & 63.89 MHz            &  63.96 MHz    \\ \hline
\multicolumn{1}{|l|}{\multirow{5}{*}{b}}& Slices                    & 25031 (11\%)         & 172219 (79\%)        &  136675 (63\%)\\ \cline{2-5} 
\multicolumn{1}{|l|}{}                  & DSP Blocks                & 0     (0\%)          & 0     (0\%)          &  0     (0\%)  \\ \cline{2-5} 
\multicolumn{1}{|l|}{}                  & LUTs as Memory            & 252   (1\%)          & 1458  (2\%)          &  1552  (1\%)  \\ \cline{2-5}
\multicolumn{1}{|l|}{}                  & Frequency                 & 59.37 MHz            & 54.17 MHz            &  54.49 MHz    \\ \hline
\end{tabular}\label{res:fit2}
\end{table}
\normalsize

\begin{table}[!ht]
\centering
\caption{Comparison to state-of-the-art implementations}
\label{res:compare}
\begin{tabular}{c|c|c|c|c|}
\cline{2-5}
                                            & Publication                                                                    & Workload & Throughput & Platform     \\ \hline
\multicolumn{1}{|c|}{\multirow{7}{*}{FPGA}} & \multirow{3}{*}{Haddoc2}                                                       & 3.8 Mop           & 318.48 Gop/s\footnotemark[1]            & Cyclone V    \\ \cline{3-5} 
\multicolumn{1}{|c|}{}                      &                                                                                & 24  Mop           & 515.78 Gop/s\footnotemark[1]            & Cyclone V    \\ \cline{3-5} 
\multicolumn{1}{|c|}{}                      &                                                                                & 24.8 Mop           & 437.30 Gop/s\footnotemark[1]            & Zynq XC706   \\ \cline{2-5} 
\multicolumn{1}{|c|}{}                      & \multirow{2}{*}{\begin{tabular}[c]{@{}c@{}}fpgaConvNet\\ \cite{Venieris2016}\end{tabular}} & 3.8 Mop           & 185.81 Gop/s\footnotemark[1]            & Zynq XC706   \\ \cline{3-5}
\multicolumn{1}{|c|}{}                      &                                                                                & 24.8 Mop          & 166.16 Gop/s\footnotemark[1]            & Zynq XC706   \\ \cline{2-5} 
\multicolumn{1}{|c|}{}                      & Qiu \textit{et al.}~\cite{Qiu2016}                                                         & 30.76 Gop          & 187.80 Gop/s\footnotemark[1]                 & \added{Zynq ZC706} \\ \cline{2-5}
\multicolumn{1}{|c|}{}                      & FINN~\cite{Umuroglu2017}                                                         & 112.5 Mop          & 2500 Gop/s\footnotemark[1]                 & Zynq ZC706  \\ \hline
\multicolumn{1}{|c|}{GPU}                   & CudNN R3                                                                       & 1333 Mop         & 6343 Gop/s               & Titan X      \\ \hline
\multicolumn{1}{|c|}{\multirow{3}{*}{ASIC}} & \multirow{2}{*}{\begin{tabular}[c]{@{}c@{}}Yoda NN\\ \cite{andri2016}\end{tabular}}    & 24.8 Mop          & 525.4 Gop/s               & UMC 65 nm    \\ \cline{3-5} 
\multicolumn{1}{|c|}{}                      &                                                                                & 23.4 Mop          & 454.4 Gop/s           & UMC 65 nm    \\ \cline{2-5} 
\multicolumn{1}{|c|}{}                      & NeuFlow~\cite{Farabet2012}                     & 350 Mop           & 1280 Gop/s                & IBM 45nm SOI \\ \hline
\end{tabular}
\end{table}

\section{Conclusion and Future work}\label{sec:concl}

This paper has investigated the feasibility of \emph{direct hardware mapping (DHM)} for the implementation of FPGA-based \ac{cnn} accelerators. We have demonstrated that current embedded \acp{fpga} provide enough hardware resources to support this approach. To demonstrate DHM, the \textsc{Haddoc2} tool has been introduced and used to automatically generate platform-independent \ac{cnn} hardware accelerators from high level CNN descriptions. Tactics are presented for optimizing the area and resource utilization of arithmetic blocks. DHM opens new opportunities in terms of hardware implementations of CNNs and can be extended to ASIC technologies as well as Binary Neural Networks.
\small
\bibliographystyle{unsrt}
\bibliography{Mendeley}
\end{document}